# Human-in-the-loop online just-in-time software defect prediction

Xutong Liu, Yufei Zhou, Yutian Tang, Junyan Qian, Yuming Zhou

**Abstract**—Online Just-In-Time Software Defect Prediction (O-JIT-SDP) uses an online model to predict whether a new software change will introduce a bug or not. However, existing studies neglect the interaction of Software Quality Assurance (SQA) staff with the model, which may miss the opportunity to improve the prediction accuracy through the feedback from SQA staff. To tackle this problem, we propose Human-In-The-Loop (HITL) O-JIT-SDP that integrates feedback from SQA staff to enhance the prediction process. Furthermore, we introduce a performance evaluation framework that utilizes a k-fold distributed bootstrap method along with the Wilcoxon signed-rank test. This framework facilitates thorough pairwise comparisons of alternative classification algorithms using a prequential evaluation approach. Our proposal enables continuous statistical testing throughout the prequential process, empowering developers to make real-time decisions based on robust statistical evidence. Through experimentation across 10 GitHub projects, we demonstrate that our evaluation framework enhances the credibility of model evaluation, and the incorporation of HITL feedback elevates the prediction performance of online JIT-SDP models. These advancements hold the potential to significantly enhance the value of O-JIT-SDP for industrial applications.

—————————— ◆ ——————————

## 1 INTRODUCTION

Just-In-Time Software Defect Prediction (JIT-SDP) predicts whether a commit is likely to introduce a bug or not by analyzing distributions of historical software changes [1, 2]. JIT-SDP offers several advantages over traditional component-level SDP. First, it enables the detection of defects at an earlier stage in the development process by performing the prediction at the time of check-in. Second, it allows for inspection of significantly fewer SLOC (source lines of code) to identify defects since it is a fine-grained defect prediction technique. Third, it makes the process of assigning developers to inspect the potentially defective commit less prone to errors. This is because the developer who made a particular commit is known, and developers can be assigned to inspect code commits that are most relevant to their area of expertise.

Recent studies have treated JIT-SDP as an online learning (also known as a stream learning) problem [3-6], which means that the model is incrementally tested and updated as new instances become available. Specifically, every time a new commit is incoming, it is first used as a test instance to obtain its predicted label. Then, after a certain waiting time, its observed label will be obtained. Based on the predicted and observed labels, the prediction performance of the online classifier (an instance of an online classification algorithm) is evaluated. After that, the commit is used as a training instance to update the online classifier. Researchers have explored various aspects of the above "test-then-train" O-JIT-SDP, including the "class imbalance evolution" phenomenon [3], the impact of cross-project training data on model performance [4], the challenge and mitigation of concept drift [5], the continuous evaluation procedure for O-JIT-SDP models in practice [6], cross-project O-JIT-SDP [7], using offline base learners in an online JIT-SDP scenario [8]. Undoubtedly, these studies have deepened our understanding of O-JIT-SDP and enhanced their potential for industrial applications in the future.

However, there are two important problems that remain to be resolved. The first problem is how to ensure reliability of the observed labels. Currently, "waiting-time window + Bug-Fixing Commit (BFC)" is the mainstream approach to acquiring the observed labels. Specifically, if a BFC $f$ is committed to fix a previous commit $u$ within the waiting-time window, then $u$ will be labeled as "bug-inducing" at the time that $f$ is committed. Otherwise, $u$ will be labeled as "clean" when the waiting time is exceeded. The benefit of this approach is that observed labels of commits are obtained fully automatically, no extra human effort is required to label the commit. But, such a labeling mechanism is inherently associated to noise in the observed labels. Considering a BFC that fixes commit $u$ arrives after the waiting time window, then a positive (i.e., actually defective) commit $u$ will be wrongly labeled as negative (i.e., clean). In a recent study [6], the default waiting-time window for BFC is set at 15 days. The reason why the waiting-time window cannot be set as long as we want to decrease the label noise is that a training instance may have been outdated after a very long waiting time and hence is not suitable for updating the online classifier (known as "concept drift" [3]). Considering the fact that it

• Xutong Liu and Yuming Zhou are with the State Key Laboratory for Novel Software Technology, Nanjing University.
• Yufei Zhou and Junyan Qian are with the School of Computer Science and Engineering & School of Software, Guanxi Normal University.
• Yutian Tang is with the School of Computing Science, University of Glasgow.

Please note that all acknowledgments should be placed at the end of the paper, before the bibliography (**note that corresponding authorship is not noted in affiliation box, but in acknowledgment section**).

can take months or even years for a bug-inducing commit to be labeled as bug-inducing by a BFC [3] (known as verification latency), this raises a serious concern about the quality of the observed labels, which may affect the correct evaluation and actual performance of O-JIT-SDP. The second problem is that there is a lack of support to continuous on-demand statistical tests over alternative algorithms. Without this essential support, developers face significant challenges, if not outright impossibilities, when attempting to make real-time decisions grounded in robust statistical evidence. This hampers their ability to accurately assess the practical efficacy of an online algorithm and to determine whether one algorithm statistically outperforms another. Such support is critical for identifying scenarios wherein the current model's performance falls short, and in determining whether there exists a need to improve or replace it with a more proficient alternative.

To tackle the above problems, we propose human-in-the-loop (HITL) O-JIT-SDP and the corresponding performance evaluation framework. In real development, when SQA staff use online JIT-SDP, their commits are the inputs of the prediction system. For each given input commit, the prediction system will predict whether it is a bug-inducing commit. For those commits predicted as bug-inducing, SQA staff will inspect them and determine whether they are bug-inducing or not. Clearly, such information can bolster the dependability of observed labels associated with commits. In this study, we hence propose HITL O-JIT-SDP that integrates label feedback from SQA staff into the prediction loop to enhance the accuracy of the prediction process. In other words, instead of separating SQA staff from online JIT-SDP like previous studies have done, we view them as a component of the process. That is to say, for instances that are predicted as defective, HITL O-JIT-SDP system gets their observed labels right after SQA staff's feedback. While for instances that are predicted as clean, HITL O-JIT-SDP system gets their observed labels in the same way with non-HITL O-JIT-SDP, aka, "waiting-time window + BFC". In this way, we can utilize the interaction between SQA staff and the online JIT-SDP system to enhance the reliability of the observed labels of commits. Furthermore, we propose a framework for evaluating the performance of O-JIT-SDP that addresses two key issues with existing methods. Specifically, in order to compute the significance of performance differences among multiple classifiers trained from online classification algorithms, existing work [3-5] adopted the approach of repeating experiments on each data stream multiple times to obtain multiple observations. However, this approach has two limitations. First, it depends on the randomness of the evaluated algorithm, which means that many commonly used online classification algorithms like Hoeffding Tree [9] cannot produce different performances through repeated execution. Second, this approach can be computationally expensive and time-consuming, especially when the number of compared algorithms is large or the dataset complexity is high. In light of the above limitations, we propose an alternative approach to calculate statistical tests in O-JIT-SDP. Specifically, our framework suggests using k-fold distributed bootstrap method to generate multiple observations

(aka., classifiers) for an online algorithm on a single execution. For online algorithms that do not incorporate randomness, using k-fold distributed validation is both feasible and effective. Furthermore, the Wilcoxon signed-rank statistical tests we use in our framework have reasonable Type I and Type II errors according to our experimental results on O-JIT-SDP datasets. By using our framework, developers can monitor robust statistical test results among alternative online algorithms in real time.

In sum, the contributions of this paper are as follows:

- We present an innovative idea termed "human-in-the-loop online JIT-SDP." By incorporating label feedback from SQA staff into the prediction loop, we enhance the reliability of observed labels for commits. To our knowledge, our proposed HITL online JIT-SDP model is the first of its kind. Furthermore, we provide a performance evaluation framework that equips developers to make well-informed, real-time decisions supported by robust statistical evidence.

- We undertake an experiment to investigate the effectiveness of augmenting online JIT-SDP with human-in-the-loop feedback and a rigorous model evaluation process. The experimental results, based on 10 diverse GitHub projects, illustrate that our evaluation framework substantially improves the reliability of model assessment, while the human-in-the-loop approach elevates the predictive performance of online JIT-SDP models. These enhancements hold the potential to greatly magnify the applicability of online JIT-SDP within authentic industrial contexts.

- We have made our HITL online JIT-SDP model and performance evaluation framework publicly available as an open-source implementation [10]. This implementation is available for use, replication, and improvement by both practitioners and researchers. Our tool is built as an extension of the widely-used online learning platform MOA (Massive Online Analysis) [11], we also release our extended MOA [12], allowing for easy and flexible integration into other tasks.

The structure of this paper is as follows. Section 2 provides an overview of the fundamental concepts. Section 3 introduces the HITL O-JIT-SDP model we have devised. Section 4 describes our proposed performance evaluation framework. Section 5 details our experimental setup, while Section 6 presents the experimental results. The discussion of our findings is presented in Section 7. Section 8 discusses related work. Finally, in Section 9, we conclude the paper and outline directions for future work.

## 2   PRELIMINARIES

In this section, we provide the fundamental concepts that form the basis of our study, including the "test-then-training" approach in online learning and the ground-truth labels and observed labels in O-JIT-SDP.

### 2.1 "Test-then-training" online learning

"Test-then-training" is a commonly used online learning approach (also known as "prequential" online learning) [13]. In this approach, a newly arrived instance is first used as a test instance to evaluate the performance of the current classifier and then used as a training instance to incrementally update the online classifier. This approach is particularly useful in situations where new data arrives continuously and the classifier needs to be updated on the fly to adapt to the changing data distribution.

During "test-then-training" online learning, the online classifier is updated continuously with the arrival of new instances. Therefore, forgetting mechanisms are commonly used when calculating the performance of classifiers to reflect their recent performances. The most commonly used forgetting mechanisms are sliding windows and fading factors [14, 15]. The sliding window mechanism calculates the performance of a classifier with a sliding window overtime of the most recent test instances. Its disadvantage is the requirement of additional memory to store the historical data. The size of the sliding window directly affects the amount of memory required, which can become a problem when dealing with large amounts of data. In contrast, the fading factor mechanism updates the performance of a classifier by a fading factor $\alpha$. Specifically, the performance of a classifier at time $i$ with fading factor $\alpha$ is as follows [15]:

$$M_i = \frac{S_\alpha(i)}{N_\alpha(i)}$$

with

$$S_\alpha(i) = x_i + \alpha \times S_\alpha(i-1)$$
$$N_\alpha(i) = 1 + \alpha \times N_\alpha(i-1)$$

where the fading sum $S_\alpha(i)$ is the *fading sum* of observations computed at time $i$ and $N_\alpha(i)$ is the corresponding *fading increment* ($S_\alpha(1) = x_1$ and $N_\alpha(1) = 1$). Note that a larger fading factor $\alpha$ means more weight to memories of past performance and a smoother model performance curve over time. As can be seen, compared with the sliding window mechanism, fading factor mechanism is memoryless and can adapt to changes in the data stream without storing the historical data. Therefore, previous studies on online learning suggest that it is advisable to calculate evaluation metrics using fading factors [15].

In recent O-JIT-SDP studies [3-8], the "test-then-training" paradigm with fading factors was used to evaluate model performance (the fading factor was often $\alpha = 0.99$). In our study, we follow these studies to adopt the same approach to update and evaluate classifiers.

## 2.2 Ground-truth and observed labels

Assigning labels to commits is a crucial aspect of O-JIT-SDP tasks. In an ideal scenario, we would have access to the "ground truth" labels of commits, which would allow the online classifier to be trained and tested without any label noise. However, in practice, there is no reliable method that can determine the actual "ground truth" labels of software changes.

As a compromise, the labels used in O-JIT-SDP are known as "observed labels." These labels are obtained from available information from the investigated project, such as BFC [3]. Current O-JIT-SDP studies use an approach called "waiting-time window + BFC" to generate observed labels for commits. Specifically, a waiting time window $W$ is set for new commits, for a commit $u$ produced at time $U$, if a bug-fixing commit $f$ appears at $T$ and $T < W + U$, and this bug-fixing commit $f$ fix software changes in commit $u$, then $u$ is labeled as bug-inducing at time $T$, otherwise, $u$ is labeled as clean at $W + U$.

In practice, there is a trade-off between label quality and concept drift when using the "waiting-time window + BFC" approach. A longer time window $W$ can reduce the number of bug-inducing commits predicted as clean, which helps reduce label noise. However, using a longer time window can also lead to concept drift, where the classifier's performance may degrade due to being trained on outdated commits. Therefore, it is not optimal to choose an excessively long waiting window to minimize label noise at the expense of increased concept drift. A balance must be struck to effectively manage both concerns. Cabral et al. found that statistics on ten open-source projects indicated that only half of the defect-inducing commits would have correct labels for training before the waiting time of 90 days ended [3]. This highlights the high level of label noise (about 50%) for bug-inducing commits and the potential for concept drift to affect model performance after three months. In a more recent study [6], Song et al. found a significant difference in the noise level of labels between 15 days BFC waiting time and 90 days BFC waiting time, and they found a BFC waiting time of 90 days caused a median drop of 3.73% in the validity compared with that of 15 days. Therefore, they selected 15 days as the default BFC waiting time in their experiments.

# 3 HUMAN-IN-THE-LOOP ONLINE JUST-IN-TIME SOFTWARE DEFECT PREDICTION

In this section, we propose HITL O-JIT-SDP. We begin by introducing the motivation for human-in-the-loop integration. Then, we present the architecture of the HITL O-JIT-SDP system. Finally, we describe the involved real-time streaming data in human-in-the-loop.

## 3.1 Motivation for integration

As mentioned previously, the current method of acquiring observed labels in O-JIT-SDP studies faces challenges related to label noise and concept drift. Therefore, a more effective approach is necessary to obtain high-quality observed labels for commits in the data stream. To tackle this issue, we propose the integration of feedback from SQA staff into the O-JIT-SDP loop, enabling the acquisition of superior observed labels for a subset of commits. In the realm of O-JIT-SDP, SQA staff are expected to engage with the O-JIT-SDP system as part of their daily routine. This involvement entails inspecting defective commits predicted by the system and providing feedback on the inspection outcomes. However, to the best of our knowledge, no prior study in the field of O-JIT-SDP has accounted for the interaction between humans and online learning systems. Consequently, existing O-JIT-SDP studies have overlooked a crucial phenomenon: there are two primary sources of observed labels in real-world scenarios, namely, delayed observed labels obtained through a "waiting time window + BFC" mechanism, and observed labels acquired from SQA staff.

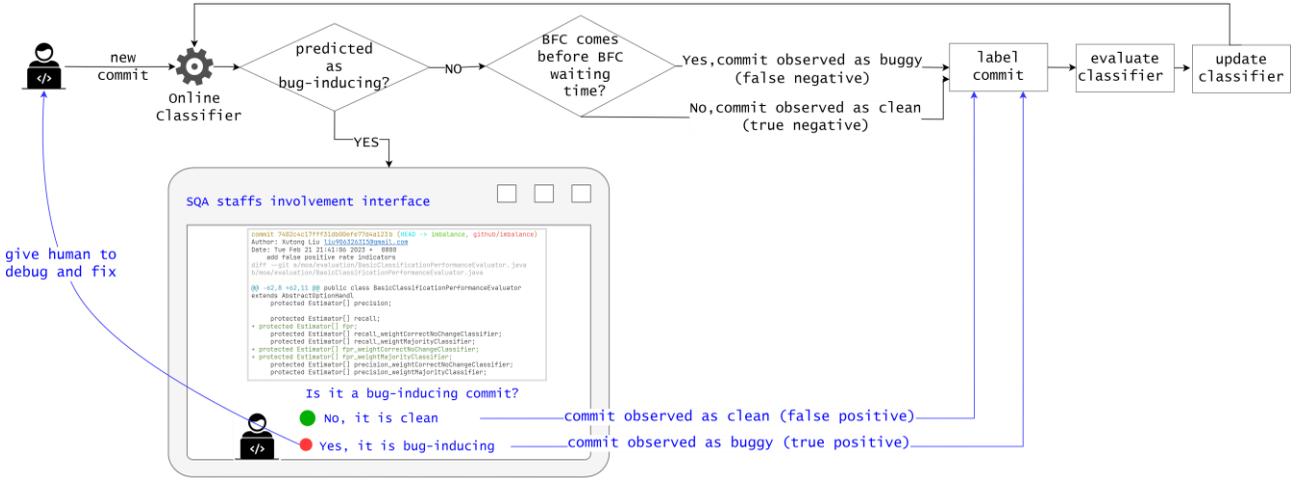

Fig. 1. The high-level architecture of HITL O-JIT-SDP

Specifically, we can take the feedback from SQA staff into the loop of O-JIT-SDP, to obtain high-quality observed labels for commits. In real-world scenarios, SQA staff treat commits predicted as defective and commits predicted as clean differently when interacting with online SDP systems. On the one hand, for commits predicted as defective, they are sent to SQA staff for inspection. The feedback provided by SQA staff, indicating whether these commits are actually defective or clean, can be used as their observed labels. On the other hand, commits predicted as clean are omitted by SQA staff, and the "waiting time window + BFC" approach can be used to generate their observed labels, similar to the non-HITL O-JIT-SDP process. This approach offers two benefits. On the one hand, involving SQA staff enables faster labeling compared to a labeling procedure based on only "waiting time window + BFC", as quality assurance is an ongoing task in software development. Consequently, the time delay in updating classifiers with labeled commits is reduced, mitigating the issue of concept drift. On the other hand, since SQA staff possess extensive project-specific knowledge, their judgment regarding the defectiveness of software changes is considered reliable. Therefore, the observed labels provided by them are expected to be less noisy compared to those obtained through non-HITL O-JIT-SDP based on "waiting time window + BFC". In this way, the human-in-the-loop (HITL) approach can improve data quality and alleviate concept drift simultaneously by incorporating the expertise of SQA staff into the labeling process.

In the domain of artificial intelligence, the concept of "Human-in-the-loop" refers to AI systems where both human and machine performance collaboratively contribute to enhancing overall outcomes and expediting the learning process. In such systems, humans engage in continuous interaction with the machine to train, monitor, and update models, particularly when deploying AI models in real-world scenarios. The application of HITL spans various stages of the lifecycle of a learning model [16], encompassing tasks such as dataset annotation [17], edge case handling [18, 19], model training [16, 20], model predictions inspection [21], outlier monitoring [22], and model explanation [23]. Currently, HITL has been successfully employed in diverse fields, including natural language process [24, 25], computer vision [19, 20], health care [26], security system [27], finance [28], and automatic program repair [29]. The emergence of HITL AI underscores the significance of human interaction in the development and advancement of AI technology, which is also applicable to O-JIT-SDP.

### 3.2 High-level architecture

Fig. 1 shows the architecture of HITL O-JIT-SDP, where human involvement is denoted by the blue elements. Whenever a new software change is committed, the online classifier generates a prediction using the features of the commit. If the commit is predicted as a bug-inducing change, it is forwarded to SQA staff for inspection, allowing humans to determine its observed label within a short time delay. Conversely, if the prediction designates the commit as clean, the observed label is assigned after a more extended delay (e.g., 15 days, following the default waiting time established in a previous study [6]). This delay is implemented using the "waiting time window + BFC" methodology. Once the observed label is obtained, it is used to evaluate this classifier and subsequently update this classifier accordingly.

As depicted in Fig. 1, commits predicted to be bug-inducing are transmitted to SQA staff through an interactive interface. Once a commit is identified as bug-inducing, it triggers the creation of a bug report that is then forwarded to developers. This leads to the initiation of a new bug-fixing commit, which becomes part of the iterative process. In this scenario, the commit carries an observed label of "buggy," representing a true positive outcome. Conversely, if the commit is deemed non-bug-inducing, the observed label is marked as "clean," indicating a false positive. Unlike previous models, such as O-JIT-SDP [3-6, 30], which typically treated this process as a non-HITL procedure devoid of SQA involvement (as shown in the blue segment of Fig. 1), our approach presents a more pragmatic scenario. We integrate the active engagement of SQA staff, who inspect suspicious software commits and provide feedback to the online classifier, thereby reflecting a more realistic operational environment.

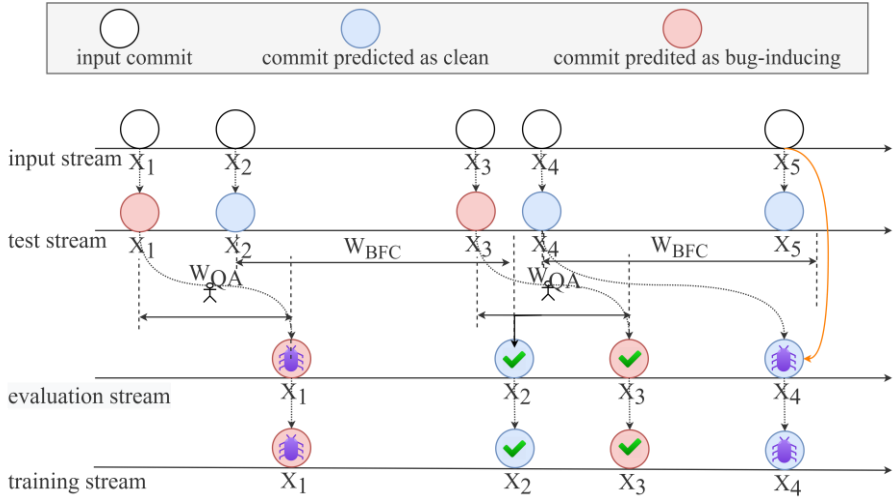

Fig. 2. Streaming data in HITL O-JIT-SDP

### 3.3 Involved streaming data

Fig. 2 provides an overview of the streaming data in HITL O-JIT-SDP, highlighting the generation and utilization of true positive, false positive, true negative, and false negative instances (i.e., commits) within the system. It elucidates how these instances are produced and employed throughout the HITL O-JIT-SDP process.

The entire HITL O-JIT-SDP process involves four types of streaming data: input stream, test stream, evaluation stream, and training stream. When a new software commit is created, it enters the input stream. Depending on the employed sampling strategy, commits in the input stream will either enter the test stream or be discarded. In Fig. 2, for simplicity, we assume no sampling strategy is applied, meaning all commits from the input stream enter the test stream. Within the test stream, the online classifier built from an online classification algorithm provides predicted labels indicating the likelihood of defects in the commits. These commits are then separated into two queues to await their observed labels.

- *positiveQueue*. Commits predicted as bug-inducing (represented by red, e.g., $X_1$ and $X_3$) are placed in the *positiveQueue* queue. For each commit in *positiveQueue*, it will be popped and sent to SQA staff for code inspection. The length of time it takes for the inspection process until obtaining the observed label is denoted as $W_{QA}$. As a result, the commit will be either confirmed as bug-inducing (true positive, e.g., $X_1$) or clean (false positive, e.g., $X_3$).

- *negativeQueue*. Commits predicted as clean (represented by blue, such as $X_2$ and $X_4$) are placed in the *negativeQueue* queue. For each commit in *negativeQueue*, if a BFC that resolves the corresponding issue appears before the waiting time $W_{BFC}$ elapses (e.g., $X_4$), it will be popped from *negativeQueue* and labeled as bug-inducing (false negative). If no BFCs that fix the commit appear until the waiting time $W_{BFC}$ (e.g.,

$X_2$) passes, it will be popped from *negativeQueue* and labeled as clean (true negative).

As can be seen, in the O-JIT-SDP process, commits in *positiveQueue* have their observed labels determined by SQA staff, while commits in *negativeQueue* have their observed labels determined via the "waiting time window + BFC" method. After obtaining the observed labels, these commits will be sent to the evaluation stream to incrementally update the classifier's performance using the "observed ground truth" labels and the previously predicted labels. After being evaluated, the commits enter the training stream, updating the online classifier using their features and the newly acquired observed labels.

In the aforementioned HITL O-JIT-SDP process, the implementation requires the setting of two parameters: $W_{BFC}$ and $W_{QA}$. Previous studies have typically used 15 days as the default value for $W_{BFC}$[6], and we adhere to this practice in our own study. In realistic scenarios, the inspection time $W_{QA}$ may vary for each commit. On the one hand, a longer $W_{QA}$ can increase the possibility of concept drift affecting the classifier, rendering the commit unsuitable for updating the classifier. On the other hand, to ensure the acquisition of dependable observed labels, SQA staff need sufficient time to inspect an incoming commit. To simplify the implementation, we treat $W_{QA}$ as a constant. To determine a suitable default value for $W_{QA}$, we examined the typical development frequency and debug frequency in 10 open-source projects (detailed in Section 5.1) to estimate the average $W_{QA}$ required for SQA staff in HITL O-JIT-SDP systems. These 10 open-source projects have been active for at least two years and are still active as of 2021. On average, these projects witness 9.1 commits per day and 2.8 bug-fixing commits (BFCs) per day, indicating a frequent occurrence of day-to-day development and code review in these active projects. Based on this analysis, we assume that commits sent to SQA staff can be inspected daily when a project is under active development and maintenance. Consequently, we set the default value of $W_{QA}$ to 7 days, allowing for large tolerance in the time consumed by code inspection. The impact of the WQA value on the performance of O-JIT-SDP will be discussed in Section 6.1.

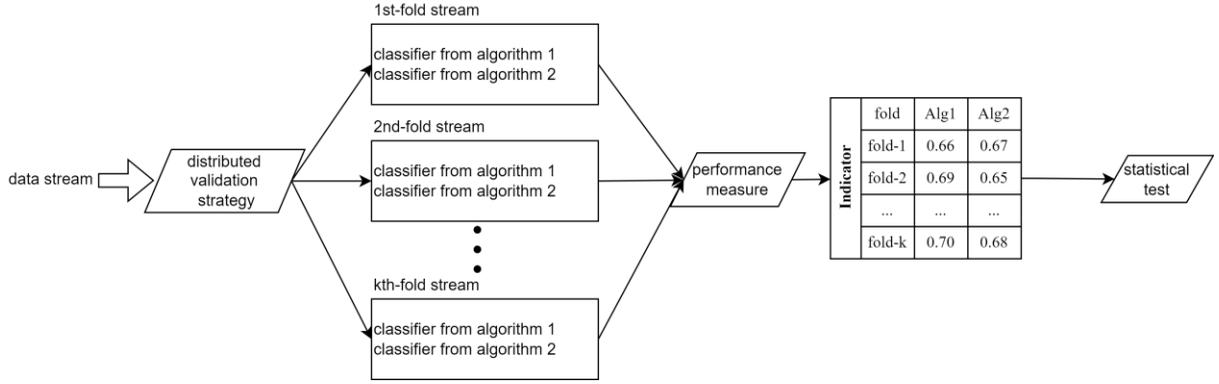

Fig. 3. The framework for online statistical performance evaluation.

## 4 ONLINE STATISTICAL PERFORMANCE EVALUATION

We present a performance evaluation framework that combines k-fold distributed validation with statistical testing. This approach facilitates pairwise comparisons between diverse classification algorithms using a prequential evaluation. Through this framework, developers gain the capability to conduct ongoing statistical testing throughout the complete prequential process. This empowers them to make well-informed, real-time decisions supported by strong statistical evidence.

Fig. 3 illustrates our framework for evaluating performance. We use a distributed validation strategy to partition the input commits into k-folds. Subsequently, we subject each alternative algorithm to testing, evaluation, and training within the k-folds partitions. This approach generates $k$ classifiers for each online classification algorithm concurrently, based on the $k$-fold partitions of the data stream. This process facilitates the generation of paired indicator values, akin to the representation in the table depicted in Fig. 3. Finally, real-time statistical tests can be executed using the generated table. We delve into the determination of the $k$-fold distributed validation strategy in Section 4.1, while Section 4.2 addresses the selection of appropriate statistical tests for analysis.

### 4.1 Online k-fold distributed validation

When comparing two different JIT-SDP models over the same commit data stream, it becomes imperative to perform rigorous statistical testing for discerning any performance disparities at specific time points. This is indispensable as it facilitates the automation of diverse decisions grounded based on robust statistical evidence. For instance, it aids in detecting scenarios where the current model's performance is lacking, and in determining whether there exists a necessity to enhance or substitute it with a more adept alternative. While Song et al. [6] introduced a prequential process that continuously evaluates the predictive performance of JIT-SDP models, the methodology for conducting on-demand, anytime statistical tests to compare alternative models remains unclear.

To address the aforementioned challenge, we adapt the k-fold distribution validation strategies proposed by Bifet et al. [13] for the context of online just-in-time defect prediction. We introduce three alternative techniques for conducting k-fold validation in the context of algorithms that learn from data streams. The core concept involves distributing data instances among k distinct versions (classifiers) of an algorithm, utilizing the following approaches:

- *K-fold distributed cross-validation.* Upon the arrival of each commit, it is used for testing in one classifier and used for training in all the others.
- *K-fold distributed split-validation.* Upon the arrival of each commit, it is used for training in one classifier and used for testing by all the others.
- *K-fold distributed bootstrap-validation.* Upon the arrival of each commit, it is used for training in each classifier according to a weight from a Poisson(1) distribution, and used for testing by all the others.

Table 1 provides an overview of the properties associated with these three k-fold distributed evaluation strategies, as proposed by Bifet et al. [13]. As can be seen, the k-fold distributed cross-validation strategy involves training each classifier with $(k$-1)/$k$ of the commits and testing with 1/$k$ of the commits. The training instances' overlap ratio between any two classifiers is $(k$-2)/$(k$-1). In contrast, k-fold distributed split-validation allocates 1/$k$ of instances for training and $(k$-1)/$k$ for testing. The training instances between any two classifiers do not overlap. Finally, k-fold distributed bootstrap-validation involves training with approximately 63.2% of commits and testing with around 36.8% of commits. The training instances' overlap ratio between any two classifiers is approximately 40%, as computed based on $(1 - P(x=0, \lambda=1))^2$.

As can be seen, Table 1 highlights the suitability of the k-fold distributed cross-validation strategy in scenarios with limited available training data. Nonetheless, this approach entails a trade-off involving a notable redundancy of training instances within each fold. On the other hand, the k-fold distributed split-validation strategy becomes preferable when ample data is at hand. This strategy's advantage lies in training each fold on distinct segments of the stream, but it is accompanied by the drawback of underutilizing the available data.

Table 1. Properties of k-fold distributed data split strategies

| Strategy | %training instances | %testing instances | %overlap ratio for any two folds |
|---|---|---|---|
| Cross-validation | $(k-1)/k$ | $1/k$ | $(k-2)/(k-1)$ |
| Split-validation | $1/k$ | $(k-1)/k$ | 0 |
| Bootstrap-validation | 63.2% | 36.8% | 40% |

Considering the analysis, we recommend the adoption of the k-fold distributed bootstrap-validation for just-in-time defect prediction due to the following reasons:

- Data stream characteristics. In real-world software development, commit streams are not as high-speed as other streams like advertising predictions. Hence, the low utilization of data for training in k-fold distributed split-validation is not well-suited for commit data streams, which may not be dense enough to support this approach effectively.

- Overlap ratio. The substantial overlap in training instances between different folds in k-fold distributed cross-validation diminishes the performance variance across folds, making statistical tests less meaningful. Moreover, it complicates the detection of performance differences among algorithms.

- Compromise approach. K-fold distributed bootstrap-validation strikes a balance between cross-validation and split-validation. It ensures a reasonable portion of training instances and a consistent overlap ratio, irrespective of the number of folds (k). This approach guarantees sufficient training data while preserving statistical reliability.

In conclusion, taking into account the characteristics of commit data streams, the limitations of alternative validation methods, and the advantages of k-fold distributed bootstrap-validation, we opt for this strategy to evaluate the prediction performance of O-JIT-SDP algorithms. This strategy presents two key advantages. First, it eliminates the necessity for evaluated classifiers to incorporate randomness. This is particularly significant for online algorithms like HoeffdingTree [9], HoeffdingAdaptiveTree [31], and HoeffdingOptionTree [32], which do not inherently involve randomness. Second, this strategy allows for incremental statistical test calculations in a single execution for input data streams. This efficiency is particularly valuable when dealing with large datasets or complex classifiers under evaluation, obviating the need for repeated executions to obtain multiple observations.

### 4.2 Statistical tests over commit stream

In the realm of online learning research [13, 33], three non-parametric statistical tests stand out: the Wilcoxon signed-rank test, the sign test, and McNemar's test. These tests serve to determine whether one classification algorithm holds a statistically significant advantage over another classification algorithm. Remarkably, all three tests are capable of providing real-time test results through k-fold distributed validation. Consequently, we identify them as prime candidates for suitable statistical tests in the context of online JIT defect prediction.

**McNemar's test.** The test statistic for McNemar's test, which is a variant of the chi-squared test, involves counting two variables for a pair of classifiers. These variables are denoted as follows: $A$ represents the number of instances that are misclassified by algorithm 1 but correctly classified by algorithm 2, while $B$ represents the number of instances that are misclassified by algorithm 2 but correctly classified by algorithm 1.

$$\chi^2 = (A - B)^2 / (A + B)$$

The test statistic in McNemar's test follows a chi-squared distribution with one degree of freedom. To determine the statistical significance, the p-value (denoted as $p$) is compared to a predetermined significance level $\alpha$, typically set at 0.05. If $p < \alpha$, we reject the null hypothesis, indicating a statistically significant difference between the two algorithms. Otherwise, given the current evidence, we cannot reject the null hypothesis, i.e., we have no evidence to show that there is a significant difference in performance between the two algorithms.

**Wilcoxon signed-rank test.** The Wilcoxon signed-rank test, unlike McNemar's test, requires calculation across multiple folds of performances (referred to as trials) for the algorithms being compared. Each fold represents a trial, and the performance on each fold is considered as an observation. To perform the test, the following test statistic using is calculated:

$$T = \min(W+, W-)$$

where $W+$ is the sum of the ranks of the positive differences (where algorithm 2 outperforms algorithm 1) and $W-$ is the sum of the ranks of the negative differences (where algorithm 1 outperforms algorithm 2). For the test statistic $T$, the corresponding p-value (denoted as $p$) can be obtained using the normal distribution.

**Sign test.** The sign test, similar to the Wilcoxon signed-rank test, is performed on multiple folds of performances for the algorithms being compared. However, unlike the Wilcoxon signed-rank test, the sign test focuses on the direction of the differences between paired observations, rather than their magnitude. It involves comparing the signs of the differences between the paired observations. First, calculate the differences between paired observations. Assign a plus sign (+) to observations where the difference is positive, a minus sign (-) to observations where the difference is negative, and exclude observations where the difference is zero. Then, count the number of plus signs and the number of minus signs. Finally, use the binomial distribution to calculate the p-value, which represents the probability of observing the given distribution of plus and minus signs by chance. The sign test is robust against outliers and does not rely on specific assumptions about the underlying distribution of the data, making it a useful non-parametric test in various scenarios.

To our knowledge, there is no prior work that has specifically applied the aforementioned non-parametric statistical tests (such as McNemar's test, Wilcoxon signed-rank test, and the sign test) to evaluate JIT-SDP under the prequential setting. In Section 5.4.2, we will design an experiment to investigate which of these tests are more suitable for comparing classification algorithms in a runtime environment. This experiment will provide valuable insights into their performance and effectiveness when applied to JIT-SDP classifiers, helping to establish a reliable and appropriate evaluation methodology for such systems.

# 5 EXPERIMENTAL SETUP

This section describes the experimental setup used to evaluate the proposed HITL O-JIT-SDP model and the corresponding evaluation framework, including our research questions, datasets, performance indicators, and the data analysis methodology.

## 5.1 Research questions

In order to validate the effectiveness of the integration of HITL into O-JIT-SDP, we first investigate the influence of SQA waiting time and BFC waiting time on the evaluation validity of HITL O-JIT-SDP.

**RQ1 (Influence of waiting time): How do SQA waiting time and BFC waiting time affect the evaluation validity of HITL O-JIT-SDP?**
Through RQ1, we aim to identify suitable values for $W_{QA}$ and $W_{BFC}$ that will be utilized in the subsequent experiments within our study.

Subsequently, we proceed to identify appropriate statistical tests that facilitate real-time, on-demand comparisons of alternative classification algorithms within a continuous stream of commit data.

**RQ2 (Suitable statistical testing): Which statistical tests are most appropriate for conducting online comparisons of alternative classification algorithms?**
Addressing RQ2 will lead us to recommend the optimal statistical test for conducting these comparisons effectively.

Finally, we culminate in a thorough comparison of prediction performance and assessment validity between HITL O-JIT-SDP and non-HITL O-JIT-SDP. The aim is to ascertain whether the incorporation of HITL into the prediction process amplifies prediction accuracy.

**RQ3 (Benefits of HITL Integration): What benefits arise from the integration of HITL in the context of JIT defect prediction?**
To address RQ3, we undertake a comparison between HITL O-JIT-SDP and non-HITL O-JIT-SDP in two aspects: (1) RQ3.1: their evaluation validity concerning the ideal O-JIT-SDP assessment approach, and (2) RQ3.2: the predictive performances of an identical online classification algorithm implemented within each framework. Through the exploration of RQ3, our goal is to comprehensively unravel the positive implications associated with the inclusion of HITL within the JIT defect prediction process.

## 5.2 Datasets

Table 2 shows 10 GitHub open-source projects used in our study. These projects were chosen randomly from a pool of projects that met the following criteria: a lifespan exceeding 2 years, containing over a total of 9,000 commits, and remaining active until at least 2021. The data acquisition from these 10 projects was facilitated by Commit Guru [34], an analytical tool proficient in extracting commit-level datasets and annotating BFCs. Commit Guru is capable of extracting 14 distinct metrics spanning five categories: diffusion, size, purpose, history, and experience. For a comprehensive understanding of these metrics, please refer to the original research conducted by Kamei et al. [2].

In order to obtain the ground truth labels for commits in our datasets, we established a link between the BFCs identified by Commit Guru and their corresponding bug-inducing commits. To ensure the accuracy of these "ground truth" labels, we allowed each commit a significant amount of time to potentially have its associated BFC identified. Similar to the approach taken by Song et al. [6], we excluded commits from the final period of the data stream (specifically, the last two years) from our datasets. This is because these commits are relatively recent, and the BFCs linked to them may not have surfaced yet or we cannot confidently determine their ground truth labels. As a result, the statistics presented in Table 2 are calculated based on data streams ranging from the initial period to the period excluding the last two years. The complete dataset can be accessed in our open-source replication kits [10].

## 5.3 Performance indicators

In our experiments, we employed three performance indicators: $R_1$, FPR (equal to 1 - $R_0$), and G-mean. $R_0$ signifies the recall value for clean commits, $R_1$ corresponds to the recall pertaining to commits that introduce bugs, and G-mean represents the geometric mean of $R_0$ and $R_1$. It is worth noting that most, if not all, of the existing O-JIT-SDP studies, chose $R_0$, $R_1$, and G-mean as their performance indicators for performance evaluation [3-6]. Although there is not a significant distinction between reporting FPR and reporting $R_0$, we specifically report FPR in our study to underscore the significance of achieving a low FPR for the effective practical implementation of O-JIT-SDP.

The performance indicators mentioned earlier are computed using a fading factor $\alpha$ of 0.99 within a 10-fold distributed bootstrap-validation framework employed in our experiments. This fading factor $\alpha$ introduces a mechanism where the impact of older instances gradually diminishes as the evaluation unfolds. This adjustment guarantees that the performance indicators accurately mirror the algorithm's effectiveness within the context of a dynamic and evolving commit data stream environment.

Table 2. Dataset statistics

| Project | Total changes (exclude last two years) | %Defect-inducing changes | Time period start | Time period end (exclude last two years) | Defects /day | Commits /day | Bug-fixing /day |
|---|---|---|---|---|---|---|---|
| brackets | 11914 | 29.4% | 2011/12/7 | 2019/3/5 | 1.3 | 4.5 | 1.2 |
| cppcheck(main) | 23033 | 36.4% | 2007/5/7 | 2021/1/31 | 1.7 | 4.6 | 1.8 |
| edx-platform(master) | 36932 | 26.6% | 2011/12/7 | 2020/11/30 | 3.0 | 11.3 | 2.7 |
| FFmpeg(master) | 91806 | 22.6% | 2000/1/1 | 2020/12/7 | 2.7 | 12.0 | 3.7 |
| gerrit(stable-2.12) | 11794 | 28.0% | 2008/10/21 | 2019/1/18 | 0.9 | 3.2 | 1.1 |
| gimp(master) | 46586 | 32.2% | 1997/1/1 | 2021/1/6 | 1.7 | 5.3 | 1.4 |
| git(master) | 43772 | 28.7% | 2005/10/10 | 2020/11/29 | 2.3 | 7.9 | 2.6 |
| mindspore(master) | 6887 | 40.6% | 2020/3/23 | 2020/11/11 | 12.0 | 29.6 | 9.7 |
| pip(main) | 8495 | 24.8% | 2008/10/15 | 2021/1/20 | 0.5 | 1.9 | 0.4 |
| vlc(master) | 86936 | 29.6% | 1999/8/8 | 2020/9/27 | 3.3 | 11.3 | 3.5 |
| average | 36815.5 | 29.9% | - | - | 2.9 | 9.1 | 2.8 |

## 5.4 Data analysis methodology

### 5.4.1 Set up for RQ1

For both RQ1 and RQ3, we need to evaluate the evaluation validity of HITL and non-HITL O-JIT-SDP. In the context of O-JIT-SDP, evaluation validity refers to the extent to which an evaluation process accurately reflects the "ideal" prediction performance of the classifier under evaluation. Fig. 4(a) depicts the evaluation process for the ideal scenario of O-JIT-SDP. In this ideal scenario, ground truth labels are immediately available as soon as commits appear, there is no waiting time for obtaining labels, and these labels are noise-free. For a classifier operating in ideal O-JIT-SDP, we represent its performance at timestamp $ts$ as $E_{ts}^{ideal}$. Fig. 4(b) depicts the evaluation process for non-HITL O-JIT-SDP. In this scenario, all observed labels are determined by a waiting time window $W_{BFC}$ and BFCs. For a classifier operating in non-HITL O-JIT-SDP, we denote its performance at timestamp $ts$ as $E_{ts}(W_{BFC})$. As described in Fig. 2, observed labels in HITL O-JIT-SDP are determined by both $W_{QA}$ and $W_{BFC}$. We denote the classifier's performance at timestamp $ts$ under HITL O-JIT-SDP as $E_{ts}(W_{QA}, W_{BFC})$. In [6], Song et al. defines the evaluation validity as:

$$V_{ts} = 1 - \left| E_{ts}^{ideal} - E_{ts}^{nonIdeal} \right|$$

Here, $E_{ts}^{nonIdeal}$ denotes $E_{ts}(W_{BFC})$ or $E_{ts}(W_{QA}, W_{BFC})$.

A larger value for $V_{ts}$ indicates a better validity of the continuous performance evaluation procedure. We use this definition of evaluation validity in our study. Specifically, the evaluation validity is calculated using G-mean. G-mean is the average value of 10-fold bootstrap validation. The default online classification algorithm used to build classifiers is HoeffdingTree.

In RQ1, we set the value of $W_{QA}$ varies among {1, 3, 7, 15, 60, 90} days with a fixed 15 days of $W_{BFC}$ (since 15 days is the default BFC waiting time in [6]) to investigate the impact of SQA waiting time $W_{QA}$ on the evaluation validity; and the value of $W_{BFC}$ varies among {1, 3, 7, 15, 60, 90} days with a fixed 15 days of $W_{QA}$ to investigate the impact of BFC waiting time $W_{BFC}$ on the evaluation validity.

### 5.4.2 Set up for RQ2

Among those commonly used statistical tests for comparing classification algorithms, which one is suitable for comparing O-JIT-SDP algorithms in run-time? Such a statistical test should have sensible Type I and Type II error when comparing O-JIT-SDP algorithms in run-time. The Type I error is incorrectly rejecting a true null hypothesis. In our case, a Type I error happens when two algorithms with identical performances are incorrectly considered to have different performances by the statistical test. The Type II error is incorrectly not rejecting a false null hypothesis. In our case, a Type II error happens when two algorithms with different performances are incorrectly considered to have identical performances by the statistic test. We run an experiment following the methodology used by Bifet et al. [13] to conduct the Type I and Type II error evaluation. Note that such an experiment requires an algorithm that possesses randomness. Therefore, we use the same algorithm that was used by Bifet et al. in [13], LeverageBagging [35], a HoeffdingTree-based online classification with randomness.

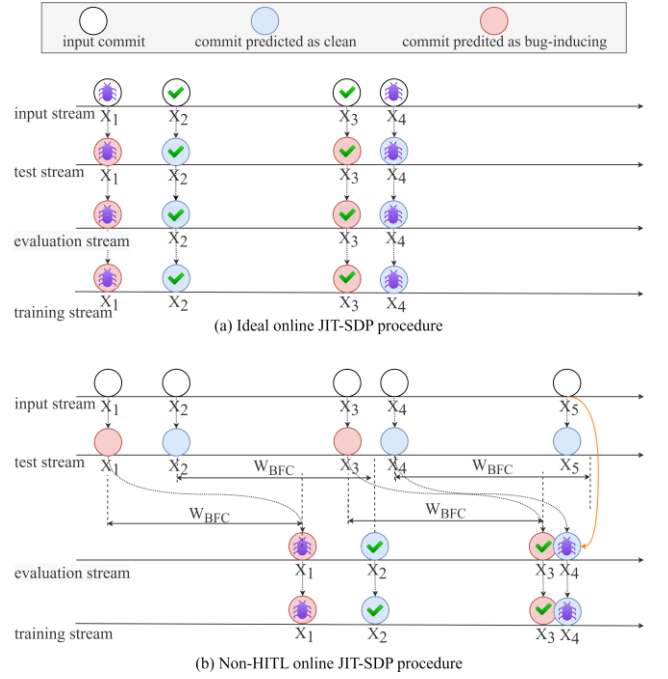

Fig. 4. Data flow in (a) ideal O-JIT-SDP and (b) non-HITL O-JIT-SDP

To check the Type I error of statistical tests, we employ distinct random seeds for identical classification algorithms. This process generates two sets of $k$-fold classifiers via k-fold distributed bootstrap-validation. Subsequently, statistical tests can be applied to paired sets of k observations. If a statistical test refutes the null hypothesis that the disparity in performance between two algorithms is negligible, and instead determines that their performance difference holds statistical significance, then the statistical test commits a Type I error.

To check the Type II error of statistical tests, we set different noise filters when building classifiers from the same classification algorithm. For example, on a noise level of 0.05, we add a noise filter that randomly reverses the predicted classification (from 0 to 1 or from 1 to 0) in a probability of 0.05. We build two sets of $k$-fold classifiers with different noise levels through $k$-fold distributed bootstrap-validation, then statistical tests can be conducted on paired $k$ observations. For two classifiers built from the same algorithm with different noise filters, if a statistical test does not reject null hypothesis, considering their performance difference is not statistically significant, then the statistical test makes a Type II error.

In RQ2, we consider classifier noise levels of {0, 0.05, 0.1}. The chosen confidence level for our statistical tests is 0.05. It is important to note that our evaluation framework allows for real-time calculation of these statistical tests. However, to optimize computational resources, we have implemented a strategy that avoids performing statistical tests on every single observation. Instead, we adopt a systematic sampling approach. At regular intervals, specifically every N/10th example in the stream, where N represents the total number of commits in the stream, we sample observations. This strategy ensures that each stream is sampled a total of 10 times, thereby generating 10 sets of k-folds observations. Subsequently, each statistical test (including the Wilcoxon signed-rank test and signed test) is carried out on these paired sets of 10 k-folds observations, resulting in a single p-value per test. It is important to highlight that McNemar's test involves a slightly distinct approach. Instead of being computed on k-fold observations as the other tests, each McNemar's test is carried out on instances within a single fold. This process is then repeated k times, yielding a comprehensive perspective on its statistical significance.

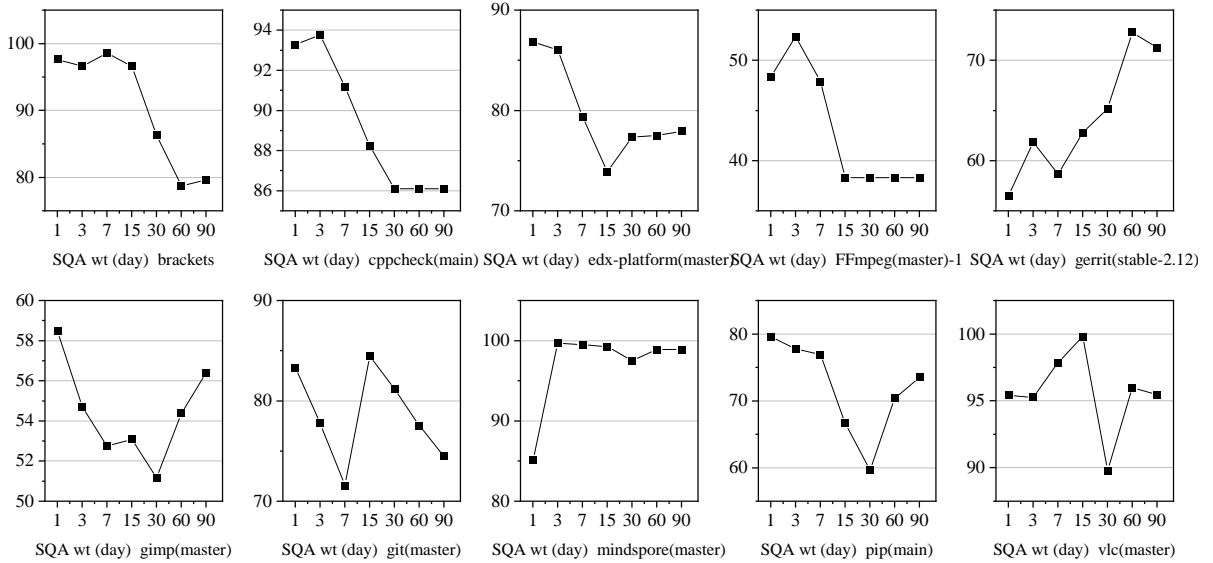

Fig. 5. Impact of SQA waiting time (x-axis) on evaluation validity of HITL O-JIT-SDP (y-axis). The BFC waiting time is fixed to 15 days.

In the context of each data stream, if the p-value resulting from the comparison between Algorithm (with random_seed1) and Algorithm (with random_seed2) is less than 0.05, it signifies the occurrence of a Type I error. Conversely, when the p-value stemming from the comparison between the classifier (with noise=0) and the classifier (with noise=0.05) surpasses 0.05, it indicates the commission of a Type II error. Similarly, if the p-value derived from the comparison between Algorithm (with noise=0) and Algorithm (with noise=0.1) exceeds 0.05, it corresponds to a Type II error. By implementing the aforementioned methodology, we iteratively apply each type of statistical test to each project, conducting this process 50 times. This enables the calculation of the Type I error rate and Type II error rate for every statistical test.

### 5.4.3 Set up for RQ3

In RQ3.1, our focus is on the comparative evaluation validity between HITL and non-HITL O-JIT-SDP. Specifically, we set equivalent waiting times for both HITL and non-HITL O-JIT-SDP scenarios, and subsequently compare their evaluation validity. This involves pitting HITL with SQA labeling (waiting time: $W_{QA}$ = 15 days, $W_{BFC}$ = 15 days) against non-HITL with BFC labeling (waiting time: $W_{BFC}$ =

15 days). The aim here is to investigate whether SQA labeling performed by HITL can enhance the evaluation validity of O-JIT-SDP, even when the time duration matches that of BFC labeling. Furthermore, we introduce an additional aspect by considering a shorter waiting time for SQA labeling (HITL with $W_{QA}$ = 7 days, $W_{BFC}$ = 15 days) against non-HITL with BFC labeling ($W_{BFC}$ = 15 days). This additional comparison aims to determine whether a swifter SQA labeling process can further enhance the evaluation validity of HITL O-JIT-SDP.

In RQ3.2, our focus shifts to the comparative analysis of prediction performance between HITL and non-HITL O-JIT-SDP scenarios. We conduct real-time monitoring of both G-mean performance and the specific statistical test highlighted in RQ2. This monitoring encompasses HITL($W_{QA}$, $W_{BFC}$) and non-HITL($W_{BFC}$) contexts, enabling us to discern variations in prediction performance between these two approaches.

## 6 EXPERIMENTAL RESULTS

In this section, we provide a thorough account of the experimental outcomes.

### 6.1 RQ1: Influence of waiting time

Two factors may affect the evaluation validity of HITL O-JIT-SDP, the SQA waiting time and BFC waiting time. Fig. 5 reports the impact of SQA waiting time on the evaluation validity of HITL O-JIT-SDP with a fixed 15 days BFC waiting time. Fig. 6 reports the impact of BFC waiting time on the evaluation validity of HITL O-JIT-SDP with a fixed 15 days SQA waiting time.

In Fig. 5, we find that the evaluation validity is higher when the SQA waiting time is generally smaller than 15 days under most projects. In most cases, the shorter the SQA waiting time, the higher the evaluation quality. However, in the real world, the SQA feedback comes after SQA staff spends time inspecting a suspicious commit.

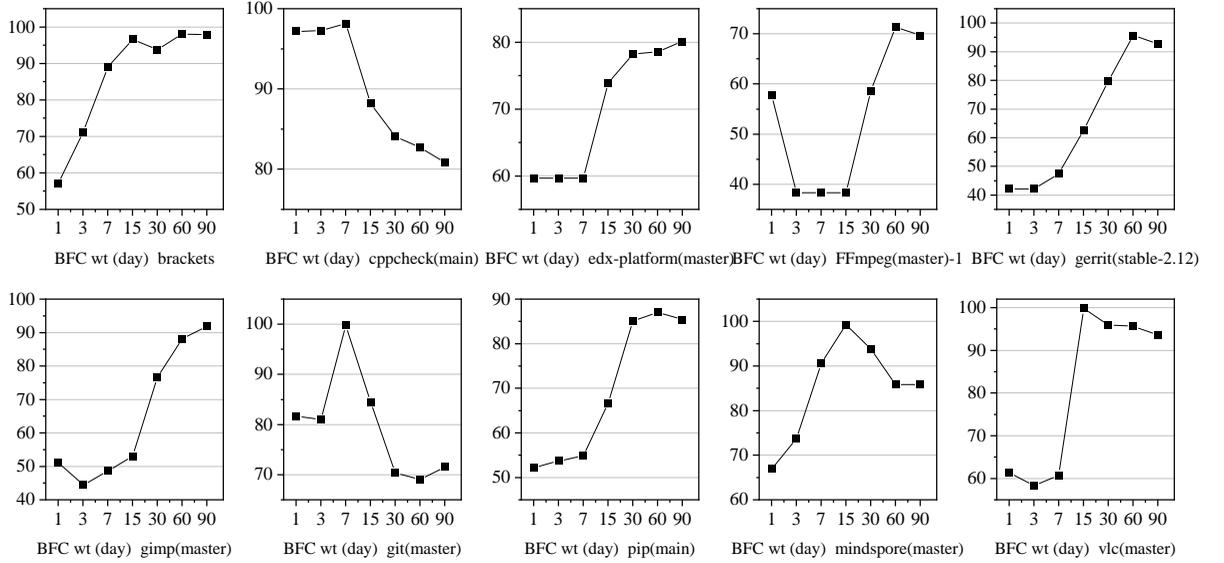

Fig. 6. Impact of BFC waiting time (x-axis) to evaluation validity of HITL O-JIT-SDP (y-axis). The SQA waiting time is fixed to 15 days.

Table 3. The ratio of null hypothesis rejection when using McNemar's test, Wilcoxon signed-rank test, and sign test under 10-fold bootstrap-validation.

| project | Mcnemar test | | | Wilcoxon test | | | Sign test | | |
|---|---|---|---|---|---|---|---|---|---|
| | No Change | Change Noise=0.05 | Change Noise=0.1 | No Change | Change Noise=0.05 | Change Noise=0.1 | No Change | Change Noise=0.05 | Change Noise=0.1 |
| brackets | 0.205 | 0.850 | 0.880 | 0.080 | 1.000 | 1.000 | 0.040 | 0.960 | 1.000 |
| cppcheck | 0.100 | 0.885 | 0.900 | 0.160 | 1.000 | 1.000 | 0.080 | 1.000 | 1.000 |
| edx-platform | 0.205 | 0.900 | 0.900 | 0.020 | 0.740 | 1.000 | 0.020 | 0.720 | 1.000 |
| FFmpeg | 0.060 | 0.900 | 0.900 | 0.060 | 0.920 | 1.000 | 0.020 | 0.560 | 0.760 |
| gerrit | 0.110 | 0.870 | 0.900 | 0.120 | 1.000 | 1.000 | 0.100 | 1.000 | 1.000 |
| gimp | 0.145 | 0.900 | 0.900 | 0.020 | 1.000 | 1.000 | 0.020 | 1.000 | 1.000 |
| git | 0.115 | 0.900 | 0.905 | 0.000 | 1.000 | 1.000 | 0.000 | 0.980 | 1.000 |
| mindspore | 0.075 | 0.805 | 0.860 | 0.040 | 1.000 | 1.000 | 0.060 | 1.000 | 1.000 |
| pip | 0.010 | 0.875 | 0.895 | 0.140 | 0.580 | 0.900 | 0.100 | 0.540 | 0.800 |
| vlc | 0.150 | 0.900 | 0.900 | 0.000 | 1.000 | 1.000 | 0.020 | 0.940 | 1.000 |
| Average | 0.118 | 0.879 | 0.894 | 0.064 | 0.924 | 0.990 | 0.046 | 0.870 | 0.956 |

*: The column labeled "No Change" pertains to Type I errors, while the remaining columns pertain to Type II errors.

Therefore, we recommend a realistic SQA waiting time of 7 days rather than shorter ones to avoid SQA people not having enough time to provide high-quality feedback.

In Fig. 6, we find an obvious difference between the evaluation validity under different BFC waiting time windows. Among those alternative waiting time windows, 15 days is a good choice for a significant portion of projects. Additionally, considering 15 days serves as the default value in the previous O-JIT-SDP research [6], we set the default BFC waiting time as 15 days.

> **Conclusion.** *SQA and BFC waiting time windows have an obvious impact on the evaluation validity in HITL O-JIT-SDP. We set the default waiting time windows as 7 days for SQA and 15 days for BFC since the evaluation validity is generally well under different projects in our experiment.*

### 6.2 RQ2: Suitable statistical testing

Table 3 presents a summary of the statistical test results. The provided values represent the proportion of rejected null hypotheses when employing McNemar's test, Wilcoxon signed-rank test, and sign test during 10-fold bootstrap validation. These tests were conducted while comparing algorithms under three conditions: no change, change with a noise level of 0.05, and change with a noise level of 0.1.

From Table 3, we have the following observations:

(1) In terms of the Type I error ratio, McNemar's test exhibits a misleading characteristic by falsely identifying non-existent differences around 11.8% of the time. Consequently, it demonstrates a pronounced inclination to overstate the disparities between algorithms. Conversely, both the Wilcoxon signed rank test and the sign test display comparable Type I error rates, at 6.4% and 4.6% respectively.

(2) In terms of Type II error ratio, the Wilcoxon signed rank test excels in distinguishing between algorithms. On average across all projects, it achieves null hypothesis rejection rates of 92.4% for change noise=0.05 and 99.0% for change noise=0.1. The sign test's average rejection rates are 87.0% for change noise=0.05 and 95.6% for change noise=0.1, while McNemar's test achieves averages of 87.9% for change noise=0.05 and 89.4% for change noise=0.1.

> **Conclusion.** *We recommend using Wilcoxon signed-rank test to compare the statistical significance of performance differences between O-JIT-SDP algorithms. McNemar's test should be avoided since its Type I error rate is large when comparing O-JIT-SDP algorithms.*

## 6.3 RQ3: Benefits of HITL Integration

### 6.3.1 Does the incorporation of HITL result in improved

*evaluation validity?*

To answer this question, we first compare the evaluation validity of HITL and non-HITL O-JIT-SDP under the same long waiting time. In Fig. 7, it becomes evident that, in the majority of instances, HITL O-JIT-SDP attains superior evaluation validity in comparison to its non-HITL counterpart. Notably, when examining scenarios such as FFmpeg(master)-1, gimp(master), git(master), pip(main), and vlc(master), the HITL evaluation approach consistently achieves heightened evaluation validity across varying waiting durations. Similarly, for other projects, the HITL evaluation process either secures higher or comparable evaluation validity across nearly all waiting periods. As outlined in section 3.1, the quality of labels can be elevated by leveraging SQA feedback to facilitate the aggregation of labeled commits. Fig. 7 reveals that the integration of SQA feedback for collecting labeled commits indeed augments evaluation validity when juxtaposed with solely relying on BFC waiting time for commit collection, even if the duration of SQA waiting time mirrors that of BFC.

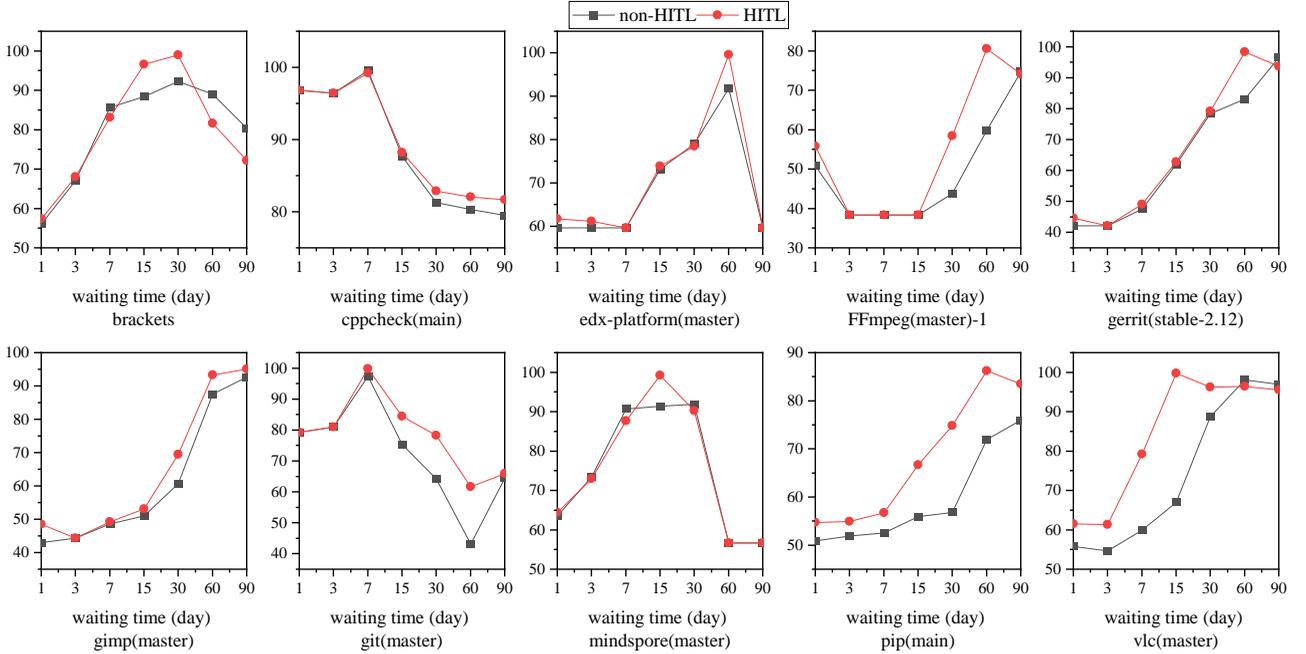

Fig. 7. Comparison of the evaluation validity (y-axis) between HITL($W_{QA}$=W, $W_{BFC}$=W) vs. non-HITL($W_{BFC}$=W). The x-axis is the value of W, which varies among {1,3,7,15,30,60,90} days.

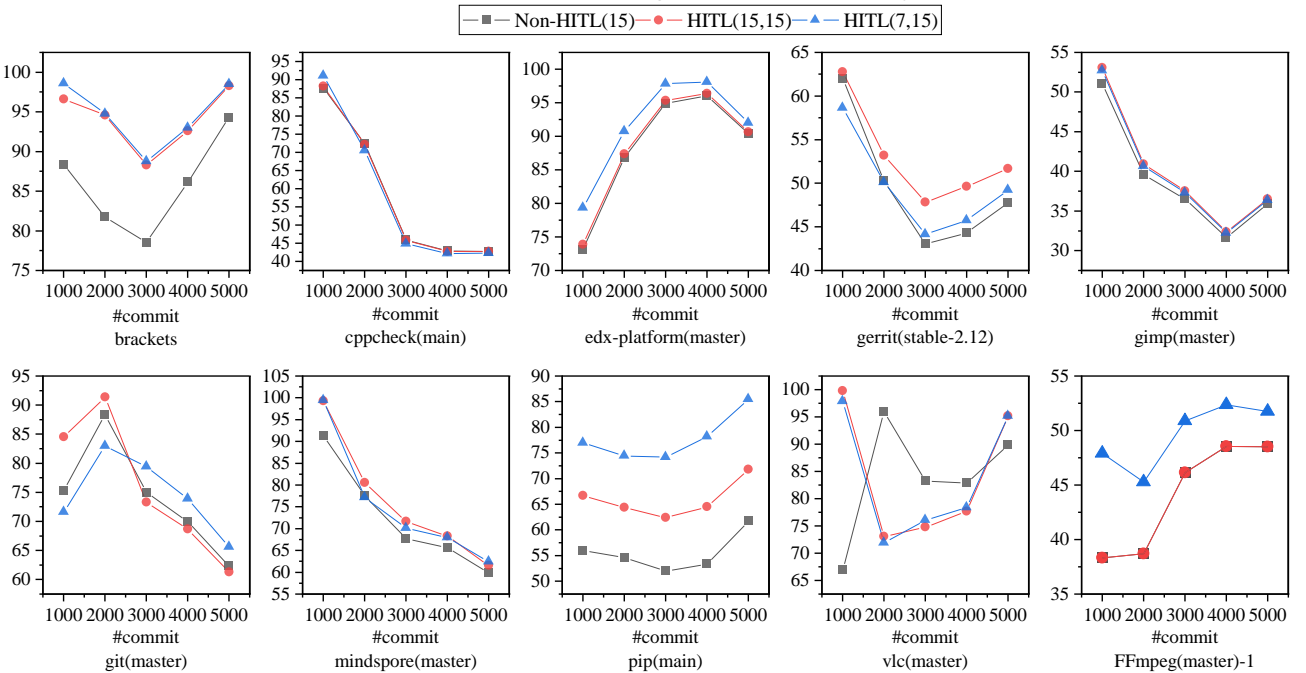

Fig. 8. Comparison of the evaluation validity (y-axis) between HITL($W_{QA}$, $W_{BFC}$) vs. non-HITL($W_{BFC}$). The x-axis is the number of top commits under evaluation.

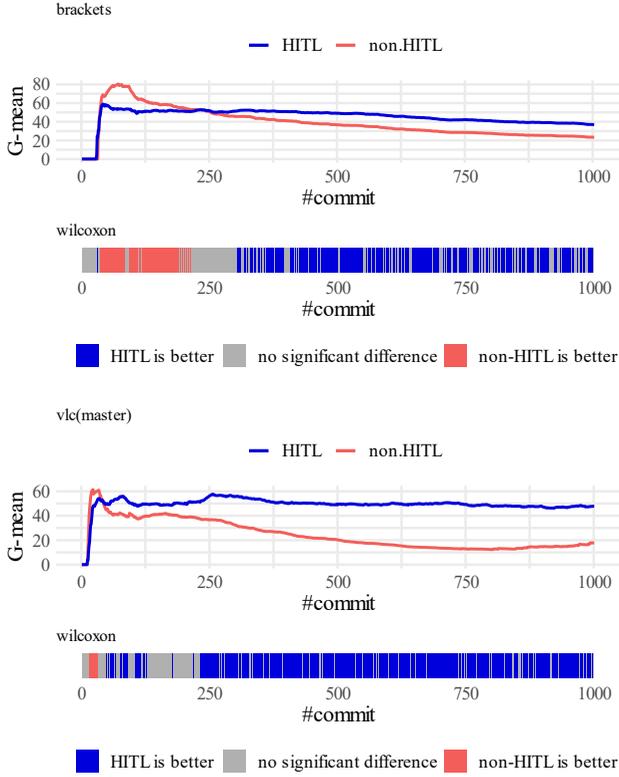

Fig. 9. Evolution of G-mean under HITL vs. non-HITL and the corresponding real-time Wilcoxon test result before the first 1000 commits. HITL denotes HITL($W_{QA}$ = 7, $W_{BFC}$ = 15) and non-HITL denotes non-HITL($W_{BFC}$ = 15).

Furthermore, another merit of the HITL O-JIT-SDP evaluation procedure is the SQA feedback can be faster than a typical BFC waiting time (15 days). Table 2 shows the average number of commits and bug-fixing commits one day are 9.1 and 2.8 separately. Based on these statistics, we assume that a suspicious commit can at least be inspected by SQA staff within one week. Therefore, we investigate whether a shorter SQA waiting time can further improve the evaluation validity compared to non-HITL O-JIT-SDP. As can be seen in Fig. 8, for most projects and most observation points (#commit), HITL($W_{QA}$=7, $W_{BFC}$=15) gets higher evaluation validity than HITL($W_{QA}$=15, $W_{BFC}$=15) and non-HITL($W_{BFC}$=15).

> **Conclusion.** *HITL not only improves O-JIT-SDP when SQA feedback is faster than BFC feedback (since it provides training commits with better label quality at a faster time) but also improves O-JIT-SDP when the SQA feedback waiting time is the same long as the BFC feedback waiting time (since it still provides training commits with better label quality).*

### 6.3.2 Does the incorporation of HITL result in improved prediction performance?

In this section, we conduct a comparative analysis of the G-mean performance between HITL and non-HITL O-JIT-SDP. Due to space limitations, we utilize two projects as illustrative examples to illustrate the real-time evolution of G-mean and the outcomes of Wilcoxon tests applied to G-mean. As depicted in Fig. 9, the performance of the same classifier, HoeffdingTree, exhibits notable superiority in terms of G-mean when operating under the HITL paradigm as compared to the non-HITL approach, particularly following a brief period of unstable cold start. The results of the Wilcoxon test affirm this observation, indicating that the classifier within the HITL O-JIT-SDP framework consistently outperforms its non-HITL counterpart after the initial cold start phase, across the majority of instances. Note that figures for all ten projects can be accessed within our replication kit [10], and they consistently demonstrate a parallel trend across the entirety of projects.

Furthermore, Fig. 9 serves as a practical demonstration of how our k-fold distributed validation-based statistical test can offer real-time oversight into the statistical performance disparities among online classifiers. This attribute holds the potential to provide invaluable assistance to developers in making well-informed decisions, encompassing the following aspects:

- *Comparative analysis of strategies*. Developers gain the ability to juxtapose diverse strategies, such as different ensemble techniques, through real-time testing. This direct comparison allows for the identification of the strategy yielding superior performance.
- *Risk management*. By continuously monitoring the statistical significance between a classifier and its baseline counterpart that developers have established, potential risks can be promptly recognized. Instances of abrupt performance deterioration can be statistically pinpointed, enabling the implementation of proactive mitigation measures.
- *Enhancement of feedback loop*. Real-time monitoring delivers instantaneous feedback to developers, thus fostering a more immediate and robust feedback loop for the enhancement of the system.

Incorporating these insights, our approach introduces a dynamic framework that empowers developers with enhanced decision-making capabilities, risk assessment, and iterative improvement of systems.

> **Conclusion.** *Under HITL, classification algorithms generally outperform non-HITL O-JIT-SDP counterparts, a difference confirmed by the Wilcoxon signed rank test.*

## 7 Discussion

In this section, we first analyze whether resampling technology can improve the performance of HITL O-JIT-SDP. Then, we condense the key takeaways for both researchers and practitioners. Finally, we delve into the potential threats that could impact the validity of our study.

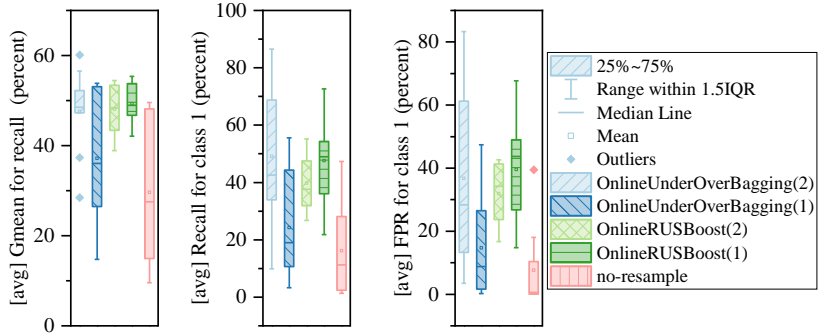

Fig. 10. Performance distribution across ten projects of Hoeffding Tree classifier with and without resampling strategies. Under the evaluation of HITL(7,15). Resampling rates of 1 or 2 adjust the positive/negative class ratio, with a rate of 2 representing a 200% positive-to-negative instances ratio.

## 7.1 Impact of data resampling

While prior studies have examined resampling strategies within the context of O-JIT-SDP [3], the validity of their assertion that "resampling strategies universally enhance O-JIT-SDP model performance" remains untested within the framework of HITL O-JIT-SDP. In this study, we employ the OnlineUnderOverBagging [36] and OnlineRUSBoost [36] resampling strategies. The former involves under-sampling the majority class and over-sampling the minority class, while the latter integrates resampling techniques as a post-processing step before each iteration of the standard AdaBoost algorithm, aligning with the techniques of OnlineUnderOverBagging. We implement these strategies using the MOA toolkit [11].

Fig. 10 illustrates the performance distributions of HoeffdingTree with and without the integration of resampling strategies. Our observations are as follows:

(1) *Enhanced $R_1$/G-mean.* It is evident that nearly all resampling strategies lead to a noticeable improvement in $R_1$ performance, regardless of the chosen resampling rate, except for OnlineUnderOverBagging(1). Notably, a significant enhancement in G-mean and $R_1$ is apparent when comparing classifiers employing resampling against the no-resample HoeffdingTree, as evident in Fig. 9.

(2) *Degraded FPR.* The majority of resampling configurations adversely affect the False Positive Rate (FPR). The classifiers utilizing resampling consistently exhibit higher FPRs in comparison to the no-resample HoeffdingTree. Furthermore, a concerning trend is observed: approximately three-quarters of resampled classifiers have a mean FPR exceeding 30% (36.7% for OnlineUnderOverBagging(2), 31.8% for OnlineRUSBoost(2), and 39.6% for OnlineRUSBoost(1)), while the mean FPR of the no-resample scenario remains considerably lower at only 7.7%.

In summary, resampling strategies tend to enhance the performance of HITL O-JIT-SDP when aiming for improvements in G-mean or $R_1$. However, it is important to note that such an enhancement often comes at the cost of an increased FPR. In the practical context of defect prediction, a high FPR is undesirable, as it can lead to an excessive number of false alarms, which developers are unwilling to address [37]. Hence, when selecting a resampling strategy for O-JIT-SDP models, careful consideration of the FPR's impact is essential.

## 7.2 Implications

Our research holds significant implications for both practitioners and researchers in the field.

### 7.2.1 For practitioners

SQA practitioners fulfill a dual role as both users of O-JIT-SDP systems and integral participants in the operational dynamics of these systems. Within the realm of O-JIT-SDP systems, the HITL variant emerges as a more pragmatic selection for SQA practitioners, primarily due to its heightened realism in contrast to the non-HITL version. Within the HITL framework, online algorithms demonstrate superior performance in comparison to their non-HITL counterparts. This trend can be attributed primarily to the increased dependability and efficiency of commit labels provided by SQA practitioners, outperforming the labels generated solely through the "waiting time window + BFC" approach. Another significant observation pertains to the adaptability of feedback timing that SQA practitioners necessitate. Our experiments underscore that a delay in feedback ranging from 1 to 15 days does not substantially compromise the validity of evaluations. This suggests that SQA practitioners can meticulously assess predicted bug-inducing commits at their own pace, thereby ensuring the delivery of feedback of the highest quality.

### 7.2.2 For researchers

Our study introduces a practical O-JIT-SDP model and an evaluation framework tailored for O-JIT-SDP. First, when proposing a new O-JIT-SDP algorithm, we advocate for its assessment within the HITL O-JIT-SDP environment to gauge its performance under realistic conditions. Second, while evaluating O-JIT-SDP algorithms, we recommend adopting $k$-fold distributed bootstrap-validation, which offers an appropriate ratio of training/testing instances and a reasonable overlap ratio of $k$-fold streams. Third, in the context of O-JIT-SDP, the Wilcoxon signed-rank test is a more fitting statistical tool compared to the sign test and McNemar's test, given its balanced consideration of Type I and Type II error rates. Fourth, while resampling strategies can enhance $R_1$ and G-mean metrics, they also elevate FPR. It is imperative to be mindful of FPR when evaluating preprocessing strategies or classification algorithms within the realm of O-JIT-SDP.

## 7.3 Threats to validity

**Construct validity.** The main threat is the potential disparity between our initial expectations regarding the role of SQA staff and their actual functions within the O-JIT-SDP framework. Uniquely, our research addresses the interaction dynamics between SQA staff and the O-JIT-SDP system, which we regard as a reflection of practical development practices. In our experimental setup, we make the assumption that SQA staff unfailingly possess accurate insights into the true labels of commits following an SQA inspection waiting period. However, this presumption might not align with reality, given that SQA staff can be prone to errors despite their considerable prior knowledge of the program. Nonetheless, our conviction remains steadfast that SQA involvement can yield swifter and more precise label feedback compared to the "BFC + waiting time" approach. In this context, our assertion that "HITL improves O-JIT-SDP" retains its validity.

**Internal validity.** The main threat is the potential limitation posed by the extent of our investigation into SQA waiting time and BFC waiting time. Given the constraints of space, we were unable to execute our experiment across a broader spectrum of SQA and BFC waiting times for RQ1. Nonetheless, we have meticulously chosen representative waiting time values, encompassing {1, 3, 7, 15, 30, 60} days, and our replication kit stands poised to facilitate deeper exploration in the future. Furthermore, it is imperative to acknowledge that our experimental foundation relies on the HoeffdingTree, an established online classification algorithm. While our conclusions are not universally generalizable to all online algorithms, it is worth noting that HoeffdingTree enjoys status as a classic and widely utilized algorithm within the context of O-JIT-SDP [3-6]. Nevertheless, the avenue remains open for further experimentation encompassing a range of algorithms through the extension of the MOA framework [12].

**External validity.** The main threat to our study is the potential lack of generalizability of our findings to other projects. We have gathered data from 10 distinct open-source projects using Commit Guru, a widely employed tool in prior O-JIT-SDP research. These projects span various domains. While we are confident that this external challenge falls within an acceptable scope, we recognize that our outcomes may not be universally applicable. To tackle this issue, we have undertaken measures to bolster the external validity of our study. By making our code and data openly accessible, we facilitate effortless replication of our research in the times ahead.

## 8 RELATED WORK

In this section, we delve into O-JIT-SDP investigations closely tied to our own study, outlining and contrasting their distinctions.

**HITL O-JIT-SDP vs. non-HITL O-JIT-SDP.** Prior studies have predominantly treated O-JIT-SDP as a non-HITL process, overlooking the potential influence of interactions between SQA staff and the O-JIT-SDP system [3-8, 30]. For instance, in [6], Song et al. introduced a predictive performance evaluation method for O-JIT-SDP within a prequential framework. Tabassum et al., in [4, 7], illustrated the benefits of fusing cross-project commit data with within-project commit data to enhance the predictive efficacy of O-JIT-SDP. In a similar vein, Cabral et al. demonstrated in [5] that PBSA (Prediction-Based Sampling Adjustment) can effectively handle various types of concept drift in O-JIT-SDP, particularly in the presence of verification latency, thereby bolstering prediction reliability. Throughout these studies, commit labels were generated solely using the "BFC + waiting time window" methodology. In contrast, our research integrates HITL principles into the O-JIT-SDP framework, resulting in commit labels that are jointly determined by both SQA staff and the "BFC + waiting time window" approach. This not only enhances the integrity of the continuous performance evaluation process but also significantly improves the predictive accuracy of just-in-time defect prediction.

**Real-time monitoring vs. repetitive experiments for statistical tests.** The majority of existing O-JIT-SDP methodologies [4-6, 30] involve running multiple iterations of each online algorithm to generate a dataset of observations. Subsequently, these observations, originating from various algorithms, are employed to conduct statistical tests. Nevertheless, the practical application of real-time statistical tests to assess alternative JIT-SDP models remains unclear. The adoption of real-time statistical tests carries substantial importance as it streamlines automated decision-making based on ongoing model comparisons. This encompasses tasks such as comparative strategy analysis, risk management, and the continuous refinement of feedback loops. To the best of our knowledge, our endeavors embody significant advancements on two fronts. First, we introduce real-time statistical test outcomes through $k$-fold distributed validation within the domain of O-JIT-SDP. Then, we provide guidance for selecting the most suitable statistical test to execute comparisons among O-JIT-SDP classifiers. This recommendation is grounded in a rigorous experiment that evaluates the Type I error rate and Type II error rate of the investigated statistical tests.

## 9 CONCLUSIONS

We propose HITL O-JIT-SDP, a realistic model of O-JIT-SDP suitable for deployment in real-world scenarios. In HITL O-JIT-SDP, commits predicted as negative and commits predicted as positive by the online classifier are treated differently. For those commits predicted as negative, their observed labels are determined based on the waiting time window and bug-fixing software changes. However, for those commits predicted as positive, their observed labels are obtained from the feedback provided by SQA staff. In previous O-JIT-SDP studies, the acquisition of observed labels for both types of commits was based on the waiting time window and bug-fixing commits. In HITL O-JIT-SDP, commits predicted as positive can obtain more reliable observed labels at a faster rate compared to non-HITL O-JIT-SDP. The experiments conducted in this study demonstrate that the same classifier performs better under HITL O-JIT-SDP than under non-HITL O-JIT-SDP.

Furthermore, we introduce an evaluation framework for O-JIT-SDP algorithms. This framework involves splitting the input data stream into $k$-fold streams using $k$-fold distributed validation strategies. Each algorithm generates $k$ instances of classifiers to learn from the $k$-fold streams individually. This process allows for the generation of $k$-paired performances, enabling a comparison between the classifiers. In order to assess the statistical significance of the differences between the compared classifiers, we recommend utilizing the Wilcoxon signed-rank test. Our experiments have demonstrated that when considering both Type I and Type II errors, the Wilcoxon signed-rank test outperforms the sign test and McNemar's test. By this evaluation framework, we can effectively compare and assess the performance of alternative algorithms.

In the future, we plan to utilize a broader array of projects to investigate the usefulness of HITL in online JIT-SDP. Furthermore, we will explore the versatility and effectiveness of HITL in various software engineering scenarios. This includes examining its applicability in predicting false alarms in static analysis warnings, forecasting crashing fault occurrences, and estimating bug severity. The objective of this investigation is to expand our knowledge of HITL's relevance in software engineering and its possible applications in various domains.